\documentclass[aps,prl,twocolumn]{revtex4}

\usepackage{epsfig}

\def\be{\begin{equation}}
\def\eea{\end{eqnarray}}
\def\ee{\end{equation}}
\def\bea{\begin{eqnarray}}
\def\ea{\end{array}}
\def\ba{\begin{array}}

\newcommand{\exval}[1]{\mbox{$\langle \, {#1}\, \rangle$}}
\newcommand{\bel}[1]{\begin{equation}\label{#1}}

\def\zzz{{\mathchoice {\hbox{$\sf\textstyle Z\kern-0.4em Z$}}
{\hbox{$\sf\scriptstyle Z\kern-0.3em Z$}}
{\hbox{$\sf\scriptscriptstyle Z\kern-0.2em Z$}}
{\hbox{$\sf\textstyle Z\kern-0.4em Z$}}}}

\begin{document}


\begin{abstract}
We investigate a recently proposed non-Markovian random walk model
characterized by loss of memories of the recent past and amnestically
induced persistence.  We report numerical and analytical results
showing the complete phase diagram, consisting of 4 phases, for this
system: (i) classical nonpersistence, (ii) classical persistence (iii)
log-periodic nonpersistence and (iv) log-periodic persistence driven
by negative feedback.  The first two phases possess continuous scale
invariance symmetry, however log-periodicity breaks this symmetry.
Instead, log-periodic motion satisfies discrete scale invariance
symmetry, with complex rather than real fractal dimensions. We find
for log-periodic persistence evidence not only of statistical but also
of geometric self-similarity.
\end{abstract}

\title{~\\Spontaneous symmetry breaking \\ in  amnestically induced
persistence}

\author{Marco Antonio Alves da Silva,$^\dagger$
G. M. Viswanathan$^{\ddagger}$, 
A. S. Ferreira$^{\ddagger}$ 
and J. C. Cressoni$^\ddagger$\\ \it
{$^\dagger$Departamento de F\'{\i}sica e Qu\'{\i}mica, FCFRP,\\
Universidade de S\~ao Paulo, 14040-903 Ribeir\~ao Preto, SP, Brazil}
\\ {\mbox{$^\ddagger$Instituto de F\'{\i}sica, Universidade Federal de
Alagoas,} Macei\'{o}--AL, 57072-970, Brazil\\}
\rm (Date: \today)
}


\maketitle

{Nonpersistent random walkers with negative feedback tend not to
repeat past behavior~\cite{elefante}, but what happens when they
forget their recent past~\cite{csv}?  Remarkably, they become
persistent for sufficiently large memory losses. This recently
reported phenomenon of amnestically induced
persistence~\cite{csv,kenkre-prl} allows log-periodic~\cite{sornette}
superdiffusion~\cite{mfs,metzler,metzler2004} driven by negative
feedback.  Its practical importance stems from the conceptual advance
of quantitatively relating, on a causal level, two otherwise
apparently unconnected phenomena: repetitive or persistent behavior on
the one hand, and recent memory loss on the other, e.g., in
Alzheimer's disease~\cite{csv}.  Precisely how does persistence depend
on recent memory loss?  Here, we answer this question and report
numerical and analytical results showing the complete phase diagram
for the problem, comprising 4 phases: (i) classical nonpersistence
(ii) classical persistence, (iii) log-periodic nonpersistence and (iv)
log-periodic persistence driven by negative feedback.  The former two
phases possess continuous scale invariance symmetry, which breaks down
in the other two.

Random walkers without memory have a mean square displacement
$\exval{x^2}$ that scales with time $t$ according to $\exval{x^2}\sim
t^{2H}$, with Hurst exponent $H=1/2$ as demanded by the Central Limit
Theorem, assuming finite moments.  Hurst exponents $H>1/2$
indicate persistence and can arise due to long-range memory.  Most
random walks with and without memory display continuous scale
invariance symmetry, i.e., continuous scale transformations by a
``zoom'' factor $\lambda$ leave the Hurst exponent unchanged: $t^{2H}
\rightarrow \lambda^{2H}t^{2H}$ as $t \rightarrow \lambda t$.

Sch\"utz and Trimper~\cite{elefante} pioneered a novel approach for
studying walks with long-range
memory~\cite{metzler,metzler2004,kenkre,kenkre2}, which we have
adapted~\cite{csv} for studying memory loss.  Consider a random walker
that starts at the origin at time $t_0=0$, with memory of the initial
$ft$ time steps of its complete history ($0 \leq f \leq 1$).  At each
time step the random walker moves either one step to the right or
left. Let $v_t=\pm 1$ represent the ``velocity'' at time $t$, such
that the position follows 
\begin{equation}
x_{t+1}=x_{t}+v _{t+1}.  \label{eq-recurr}
\end{equation}%
At time $t$, we randomly choose a previous time $1\leq t'<ft$ with
equal {\it a priori} probabilities.  The walker then chooses the
current step direction $v_{t}$ based on the value of $v_{t'}$, using
the following rule.  With probability $p$ the walker repeats the
action taken at time $t'$, and with probability $1-p$ the walker takes
a step in the opposite direction $-v_{t'}$.  Values $p>1/2$ ($p<1/2$)
generate positive (negative) feedback.  For $p$ sufficiently larger
than $p=1/2$, the behavior becomes persistent (i.e., $H>1/2$). But the
finding of persistence for $p<1/2$ and small $f$ overturned commonly
held beliefs concerning repetitive behavior and memory
loss~\cite{csv}. Very recently, Kenkre~\cite{kenkre-prl} has found an
exact solution to this problem for the behavior of the first moment,
for all $f$, and generalized it in important ways, with excellent
agreement with the numerical results over 6 orders of magnitude in
time.
  Here, we investigate how persistence depends quantitatively on
memory loss and how to characterize the important underlying symmetry
properties.

Fig. 1(a) shows values of $H(f,p)$, estimated via simulations, as a
function of the feedback parameter $p$ and the memory fraction $f$.
We choose an order parameter $2H-1$ that has positive values only in
the persistent regime.  Misleadingly, only two phases may at first
seem apparent, namely, persistent and nonpersistent.  However, the
persistent regime comprises two different phases with distinct
symmetry properties.  For $p>1/2$ we find classical persistence
satisfying continuous scale invariance symmetry. In contrast, we find
discrete scale invariance symmetry \cite{sornette} for $p<1/2$: scale
invariance holds only for discrete values of the ``zoom'' or
magnification $\lambda_k=\lambda^k$ ($k=1,2,3 \ldots$).  The
nonpersistent regime, similarly, allows for monotonically growing as
well as log-periodic solutions.  The simulation results agree well
with our analytical results discussed further below (see
Fig.~\ref{fig-main}(b)).  This spontaneous symmetry breaking indicates
a distinct phase (and eliminates the possibility of an infinite-order
phase transition) \cite{stanley}.  Mathematically, discrete scale
invariance symmetry~\cite{sornette} involves complex
fractal~\cite{mandelbrot,fractal} dimensions: for $z=a+bi$, the real
part of $t^z$ equals $t^a \cos[b\log t]$, indicating
log-periodicity~\cite{sornette}.
Fig. 1(c) parametrically plots two independent persistent
walks for $p<1/2$ ($p=0.1$ and $f=0.1$).  The walks appear not only
statistically but also geometrically self-similar.  Indeed, we find a
pattern reminiscent of a logarithmic spiral.
We estimate a value of the critical exponent $\beta=1$ from the double
log plot (Fig. 1(d)) of the order parameter versus $|p-p_c|$.



\begin{figure}
\centerline{\psfig{width=9.5 cm,clip,figure=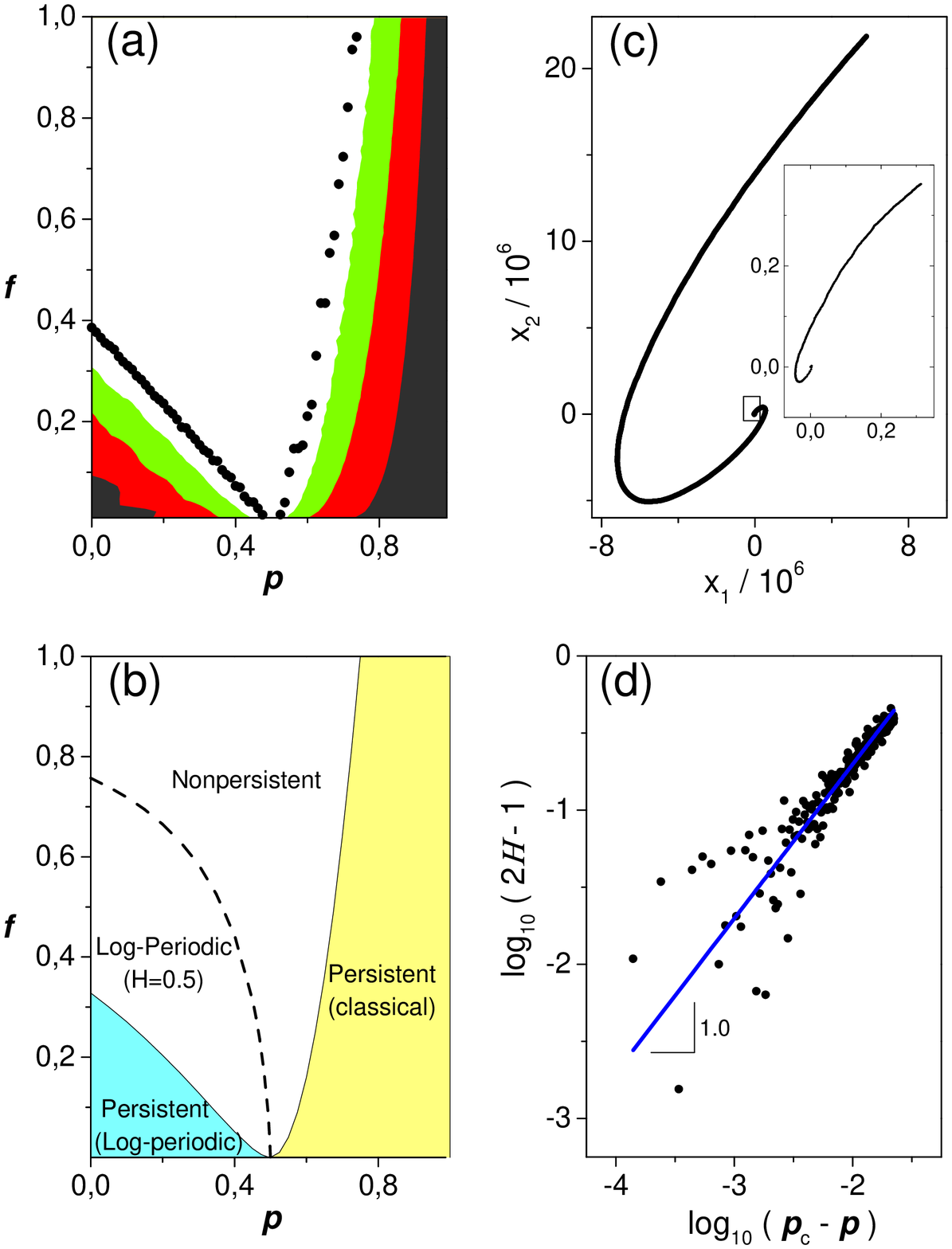}~~~}
\vspace{ -0.5 cm}
\caption{(a) Hurst exponent $H(p,f)$ estimated from simulations of
non-Markovian walks that remember only a fraction $f$ of their distant
past, with feedback parameter $p$. Color scheme: $H<0.6$ (white),
$0.6<H<0.7$ (green), $0.7<H<0.9$ (red), $H>0.9$ (black).  Full circles
show the edge where $H=1/2$.  Persistence ($H>1/2$) arises for
positive ($p>1/2$) as well as negative ($p<1/2$) feedback.  (b)
complete phase diagram showing the 4 phases, plotted according to the
exact solutions given by Eqs.~\ref{eq-fc1}, \ref{eq-fc2}. The dashed
line $f_0(p)$ delineates the threshold for log-periodicity and cleaves
the nonpersistent regime into two (Eq.~\ref{eq-kenkre}). (c)
parametric plot of positions $x_1(t)$ and $x_2(t)$ for two
realizations of log-periodic walks.  Inset shows zoom of the center
(boxed area) of the logarithmic-like spiral.  Note the geometric
self-similarity. (d) double log plot of $2H-1$ versus $p_c-p$, showing
critical behavior with critical exponent $\beta=1.$}
\label{fig-main} 
\end{figure}

\begin{figure}[t]
\centerline{\psfig{width=11cm,clip,figure=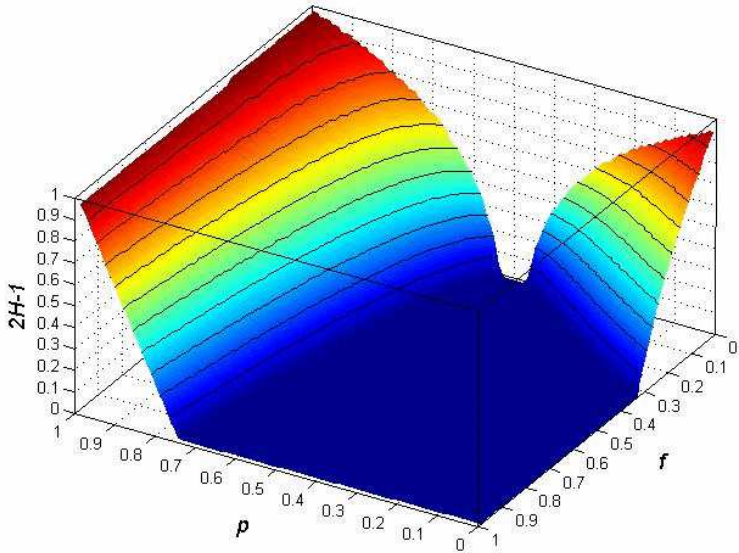}~~~}
\caption{The order parameter $2H-1$ versus $p$ and $f$, estimated
using Eqs. \ref{Hurst 1} and \ref{eq-H}.}
\label{fig-arlan}
\end{figure}

We next approach the problem analytically. We choose $v _{t^{\prime
}}=\pm 1$ and use the previously discussed recurrence relation, 
Eq.~\ref{eq-recurr}.
Let $n_{f}(t)$ e $n_{b}(t)$ denote the number of steps taken in the
forward and backward directions respectively, at time $t$ (inclusive).
The total number of steps is thus $n_{f}(t)+n_{b}(t)=t$. For full
memory, the probability to take a step in the forward direction at time 
$t+1$, for 
$t\geq 1$, is
\bigskip 
$P_{\mbox{\scriptsize eff}}^{+}(t)=\frac{n_{f}(t)}{t}p+\frac{n_{b}(t)}{t}(1-p). 
\label{prob. eff frente}
$
~Similarly, 
$
P_{\mbox{\scriptsize eff}}^{-}(t)=\frac{n_{b}(t)}{t}p+\frac{n_{f}(t)}{t}(1-p).
$
 So, the effective value expected at time $t+1$ is 
$
v _{t+1}^{e}=P_{\mbox{\scriptsize eff}}^{+}(t)-P_{\mbox{\scriptsize eff}}^{-}(t).
$
 Since 
$x_{t}=n_{f}(t)-n_{b}(t)+x_{0}$, 
we obtain 
$
v _{t+1}^{e}=\alpha \frac{x_{t}-x_{0}}{t},
$
where $\alpha =2p-1$.  We can interpret this result as a series of
experiments or walks at time $t$ having the same number of steps
forwards and backwards, giving the value $v _{t+1}^{e}$.

Now we introduce memory loss. Let the memory range be
$L=L(t)=int(ft)$, where $int(x)$ denotes the integer part of $x$,
for  $0<f\leq 1$, starting at $t=0$.
In analogy to the results above,  we obtain
$
v _{t+1}^{e}=\alpha \frac{x_{L}-x_{0}}{L}.
$

Now we study the $n$th moments of $x_{t}^{n}$.
Taking Eq. 
\ref{eq-recurr} to power $n$ we get 
$x_{t+1}^{n}=(x_{t}+v _{t+1})^{n}=\sum\limits_{i=0}^{n}\left(\! 
\begin{array}{c}
n \\ 
i%
\end{array}%
\!\right) v _{t+1}^{i}x_{t}^{n-i}$.  For all even exponents $i$ we know
$v _{t+1}^{i}=1$ and for odd exponents $v_{t+1}^{i}=v _{t+1}$. Using
the expression for $v _{t+1}^{e}$, with $x_{0}=0$, we obtain

\begin{eqnarray}
&& \langle x_{t+1}^{n}\rangle=\Delta +\exval{x_{t}^{n}}+\frac{n\alpha }{L}\exval{x_{L}x_{t}^{n-1}}+ 
\nonumber \\
&& \sum\limits_{l=1}^{s(n)}\left[    \left(\!
\begin{array}{c}
n \\ 
2l%
\end{array}%
\!\right) \exval{x_{t}^{n-2l}}+\frac{\alpha }{L}\left(\!
\begin{array}{c}
n
 \\ 
2l+1%
\end{array}%
\!\right) \exval{x_{L}x_{t}^{n-2l-1}}\right],~~~~   \label{momentos dos deslocamentos}
\end{eqnarray}%
where $\Delta =\frac{1+(-1)^{n}}{2}$ and $s(n)=\frac{n-\Delta -1}{2}$.
We have $\Delta =1$ for even $n$ and $\Delta =0$ for odd $n$. If
$s(n)<1$, then the sum vanishes.

In the asymptotic limit, we arrive at the following differential
equation for the moments, starting from Eq. \ref{momentos dos
deslocamentos}:

\begin{eqnarray}
& & \frac{d}{dt} \langle x_{t}^{n}\rangle=\Delta +\frac{n\alpha }{L}\exval{x_{L}x_{t}^{n-1}}+ 
\nonumber \\ 
& & \sum\limits_{l=1}^{s(n)}\left[ \left(\!  
\begin{array}{c}
n \\ 
2l%
\end{array}%
\!\right) \exval{x_{t}^{n-2l}}+\frac{\alpha }{L}\left(\!
\begin{array}{c}
n \\ 
2l+1%
\end{array}%
\!\right) \exval{x_{L}x_{t}^{n-2l-1}}\right].~~~~~  \label{equacao geral}
\end{eqnarray}

For the first moment ($n=1$), we obtain an equation identical to the
one obtained by Kenkre~\cite{kenkre-prl}:
\begin{equation}
\frac{d}{dt}\exval{x_{t}}=\frac{\alpha }{ft}\exval{x_{ft}}.  \label{primeiro momento}
\end{equation}
Considering an expansion of the form
\mbox{$
\exval{x_{t}}\sim \sum_i A_it^{\delta_i }\sin (B_i\ln (t)+C_i),  \label{solucao 1}
$}
we obtain a system of transcendental equations for $B$ and $\delta $:
\begin{eqnarray}
\delta &=&\alpha f^{\delta -1}\cos (B\ln f)  \label{eq trans 1} \\
B &=&\alpha f^{\delta -1}\sin (B\ln f).  \label{eq trans 2}
\end{eqnarray}

For $\alpha \geq 0$, we have $B=0$.  The system given by Eqs.  \ref{eq
trans 1} and \ref{eq trans 2} reduces to Eq.~\ref{eq-delta} below,
ruling out a log-periodic solution.  For $\delta >1/2$, we obtain
superdiffusion (see below).

For $\alpha <0$, there exists a threshold defined by a continuous set
of values $(p,f)$, with oscillating solutions.  Consider first the
case $B=0$, without oscillations. Then, Eq.~\ref{eq trans 1} becomes
\be
\delta = \alpha f ^{\delta-1}\;\; , 
\label{eq-delta}
\ee
which only has solutions for
$f>f_0(p)$. Numerically solving this equation, we find the threshold
$f_0(0)= 0.7569$ for $p=0$, in perfect agreement with the expression
\be (1-2p)\ln(1/f_0)=f_0/e \;\; ,  
\label{eq-kenkre}
\ee 
obtained by Kenkre~(personal communication, 13 July 2007) for the
onset of log-periodicity~\cite{kenkre-prl}, shown as a dashed line in
Fig~\ref{fig-main}(b).  Indeed, we have found an alternative proof of
Eq.~\ref{eq-kenkre} starting from Eq.~\ref{eq-delta}, using the
Lambert W function.
For $\delta
>1/2$, both superdiffusion and log-periodicity appear (see below).

We next study the second moment. If $n=2$ then $\Delta =1$
and $s(n)=0$. Thus, Eq. \ref{equacao geral} leads to
\begin{equation}
\frac{d}{dt}\exval{x_{t}^{2}}=1+\frac{2\alpha }{ft}\exval{x_{ft}x_{t}}.
\label{segundo momento}
\end{equation}
Using the fact that $\left| \exval{x}\right| \leq
(\exval{x^{2}})^{1/2}$, we can prove that there exists a function
$A(t)$ such that
$\exval{x_{ft}x_{t}}=A(t)(\exval{x_{ft}^{2}}\exval{x_{t}^{2}})^{1/2}$,
with $-1\leq A(t)\leq 1$.

For $\alpha \geq 0$ ($p\geq 1/2$), no oscillations appear and
$A(t\rightarrow \infty )=1$.  We can thus show that the following
transcendental relationship holds for the Hurst exponent,
asymptotically:
\begin{equation}
H=\alpha f^{H-1}.  \label{Hurst 1}
\end{equation}%
This result is identical to Eq.  \ref{eq-delta} with 
$\delta=H$. Consequently, 
the curve
\begin{equation}
f_c=16\left(p_c-\frac{1}{2}\right)^{2}, \label{transicao p > 1/2}
\label{eq-fc1}
\end{equation}%
separates the diffusive and anomalous regions for $p\geq 1/2$ in the
$(p,f)$ plane (Fig.~\ref{fig-main}(b)).  The case $f=1$ leads to
$p_c=3/4$, in agreement with ref.~\cite{elefante}.

For  $\alpha <0$, we try the expansion
\mbox{$
\exval{x_{t}^{2}}\sim \sum_i a_it^{2H_i}\sin ^{2}(b_i\ln t+c_i).$}
We assume that the dominant terms of $\exval{x_{t}}$ and
$\exval{x_{t}^{2}}$ have the same ``period'' and phase difference, so
that $b=B$ and $c=C$.  In the log-periodic region (i.e. $b\neq 0$), we
also assume that $H\geq\delta $, so that the solution follows from
Eqs.  \ref{eq trans 1} e \ref{eq trans 2}.  Indeed we conjecture that
for walks lacking subdiffusion, $\delta \geq 1/2$ implies $H=\delta$
always.  Subject to natural restrictions, the Hurst exponent must thus
satisfy
\be
H=\frac{\sqrt{\alpha^2f^{2H-2} -H^2} }
{\tan\left[ \ln(f)\sqrt{\alpha^2f^{2H-2} -H^2}\right] } \;\; 
\label{eq-H}
\ee
for $\delta\geq 1/2$ and $H=1/2$ otherwise.  The separation line of the
diffusive and anomalous phases corresponds to $H=\delta=1/2$, so the
critical line (Fig.~\ref{fig-main}(b)) satisfies
\be
2\sqrt{\frac{\alpha_c^2}
{f_c}-\frac{1}{4}}=
\tan\left[{\ln(f_c) 
~\sqrt{\frac{\alpha_c^2}{f_c}-\frac{1}{4}} }~\right]
 \;\; .
\label{eq-fc2}
\ee
We obtain the critical value of $f_c(0)=0.3284$ for the onset of
log-periodic superdiffusion, which occurs at $p=0$. Note that
$p_c=1/2, f_c=0$ represents a multicritical point.
Fig. \ref{fig-main}(b) shows the complete phase diagram, consisting of
4 phases.  Fig.~\ref{fig-arlan} shows a better view of how the order
parameter $2H-1$ depends on $p$ and $f$.

 Substituting our choice of
\exval{x_{t}^{2}} into Eq.  \ref{segundo momento}, with the definition
of $A(t)$, we obtain
\begin{equation}
A(t)=\frac{H\sin (b\ln t+c)^{2}+b\sin (b\ln t+c)\cos (b\ln t+c)}{\alpha
f^{H-1}\left| \sin (b\ln t+c)\right| \left| \sin (b\ln ft+c)\right| }.
\label{Equacao para A(t) H>1/2}
\end{equation}%
Semi-empirically, we find that $A(t)\approx \pm 1$. Specifically, we propose
\be
A(t)= \frac{ \sin (b\ln t+c) \cos (b\ln ft+c) }{ |\sin (b\ln t+c) \sin (b\ln ft+c)| }\;.
\ee
Note that 
$A(t)=1$
for $f=1$.
Applying the same reasoning for the case 
$H=1/2$, we obtain 
\bea
A(t)& =& f^{1/2} {
\big[2\alpha a\left| \sin (b\ln t+c)\right| \left| \sin (b\ln
ft+c)\right|\big]^{-1}}\nonumber \\ 
& & \times \big[ a\sin (b\ln t+c)^{2}\nonumber\\
& & \quad +2ab\sin (b\ln t+c)\cos (b\ln
t+c)-1\big] \;
.  \label{Equacao para A(t) H=0.5}
\eea
Thus, exactly on the critical line, 
we get a marginally superdiffusive solution:
%
$\exval{x^2(t)}= a t \ln t \sin^2(b \ln t + c)$.  
%
simulations suggest that higher order terms in the expansion become
important near the critical line.

We next focus on finding a suitable Fokker-Planck Equation (FPE).  Let
$Y$ denote the position of a walker.  Consider the conditional
probability at position $Y$ at time $t+1$, given
position $x_{0}$ at time $t=0$:
\bea
P(Y,t+1|x_{0},0)=&P(Y+1,t|x_{0},0)P_{\mbox{\scriptsize eff}}^{b}(t,Y+1)~~~~ \nonumber\\
+&P(Y-1,t|x_{0},0)P_{\mbox{\scriptsize eff}}^{f}(t,Y-1).~~~~
\label{Probabilidade Y}
\eea
Using the definitions of 
$n_{f}(t)$ and  $n_{b}(t)$, and 
the probabilities to go forwards or backwards,
we obtain 
\begin{equation}
P_{\mbox{\scriptsize eff}}^{\pm}(t,Y)=\frac{1}{2}\left[ 1\pm\frac{\alpha (Y-x_{0})}{t}\right] 
\label{prob. eff frente_2} \;\;. 
\end{equation}%

Substituting Eq.  \ref{prob. eff frente_2} 
in Eq. \ref{Probabilidade Y}, we get 
\begin{eqnarray}
& P&(Y,t+1|x_{0},0) \nonumber \\
&=& \frac{1}{2}\left[ 1-\frac{\alpha
(Y-x_{0}+1)}{t}\right] P(Y+1,t|x_{0},0) \nonumber \\
&+&\frac{1}{2}\left[ 1+\frac{\alpha (Y-x_{0}-1)}{t}\right]
P(Y-1,t|x_{0},0).
\label{Probabilidade Y_2}
\end{eqnarray}%
From this last equation, in the asymptotic limit for $t$ and $Y$, we
get a FPE for $x=Y$ which is identical to the one in
ref.~\cite{elefante}, as expected. 
 Similarly, we can show for memory loss that 
\begin{equation}
\frac{\partial P(x,t)}{\partial t}=\frac{1}{2}\frac{\partial ^{2}}{\partial
x^{2}}P(x,t)-\frac{\alpha }{ft}\frac{\partial }{\partial x}[x_{ft}P(x,t)],
\end{equation}%
where $x_{ft}$ denotes the position at time $ft$ that leads to
position $x$ in the future, at time $t$. 


In summary, we have uncovered the essential features of the phase
diagram for this problem, based on numerical as well as analytical
results.  We expect the phase diagram and other findings reported here
to contribute towards a better quantitative description of persistence
in diverse economic~\cite{econo}, sociological~\cite{west-soc},
ecological~\cite{turchinbook}, biological~\cite{turchinbook,mfs} and
physiological~\cite{mfs} complex systems where recent memory loss may
play a role~\cite{csv}.  Specifically, prime candidates for further
study include systems with long-range memory that show evidence of
log-periodicity and discrete scale invariance
symmetry~\cite{sornette}.  The most important result, for systems with
negative feedback, concerns the existence of a critical threshold of
memory loss for the onset of superdiffusion.  Finally, we hope that
the insights provided by this study dispel lingering doubts that
negative feedback coupled with recent memory loss can in fact cause
persistence.

\bigskip

\section*{Acknowledgements} 
We thank V. M. Kenkre and Marcelo L. Lyra for discussions and CNPq and
FAPESP for financial assistance.  



\begin{thebibliography}{mt1}











\bibitem{elefante} G. M. Sch\"utz and S. Trimper,
{Phys. Rev. E} {\bf 70}, 045101 (2004).



\bibitem{csv} J. C. Cressoni, M. A. A. da Silva, and G. M. Viswanathan,
{Phys. Rev. Lett.} 
{\bf 98}, 070603 (2007).



\bibitem{kenkre-prl} V. M. Kenkre, ``Analytic Formulation, Exact
Solutions, and Generalizations of the Elephant and the Alzheimer
Random Walks,'' arXiv:0708.0034v2 [cond-mat.stat-mech] 6 Aug 2007.


\bibitem{sornette} D. Sornette, 
{Proc. Nat. Acad. Sci.} {\bf 99}, 2522 (2002).



\bibitem{mfs} 
\emph{L\'{e}vy flights and related topics in physics}, edited by 
M. F. Shlesinger, G. M. Zaslavsky and
U. Frisch 
(Springer, Berlin, 1995).


%
%


\bibitem{metzler} R. Metzler and J. Klafter,
	 {Phys. Rep.}  \textbf{339}, 1 (2000).

\bibitem{metzler2004} R. Metzler and J. Klafter,
{J. Phys. A} {\bf 37}, R161 (2004).




\bibitem{kenkre} V. M. Kenkre, in {\it Statistical Mechanics and
Statistical Methods in Theory and Application}, edited by U. Landman
(Plenum, New York, 1977).

\bibitem{kenkre2} V. M. Kenkre, E. W.  Montroll and
M. F. Shlesinger,
{J. Stat. Phys.} {\bf 9}, 45 (1973).




\bibitem{stanley}
H. E. Stanley, {\it Introduction to Phase
Transitions and Critical Phenomena} (Oxford University Press,
Oxford and New York, 1971).


\bibitem{mandelbrot}
B. B. Mandelbrot, {\em The Fractal Geometry of
Nature} (Freeman, San Fransisco, 1982).

\bibitem{fractal}
Bunde A., and S. Havlin, 
{\it Fractals and Disordered Systems}
(Springer, Berlin, 1991).



\bibitem{econo}
R. N. Mantegna and H. E. Stanley,
{\it An
Introduction to Econophysics} (Cambridge Univ. Press, Cambridge 2000).


\bibitem{west-soc}
B. J. West, {\it Mathematical Models As a Tool for the
Social Sciences} (Taylor and Francis, London, 1980).


\bibitem{turchinbook}
%
P. Turchin, {\it Quantitative Analysis of Movements: Measuring and
Modeling Population Redistribution in Animals and Plants} (Sinauer
Associates, Sunderland, 1998).




%







\end{thebibliography}
\end{document}